\documentclass[12pt]{article}
\usepackage[dvips]{graphicx}
\usepackage{pdproc}

  \makeatletter 
  \def\@cite#1{[#1]} 
  \makeatother    
\begin{document}

\renewcommand{\thefootnote}{\fnsymbol{footnote}}

\title{
 Radiative corrections in SUSY phenomenology 
\footnote{Talk at the 12th International Conference 
on Supersymmetry and Unification of Fundamental 
Interactions (SUSY 2004), Tsukuba, Japan, June 17--23, 2004.}
}

\author{ YOUICHI YAMADA}

\address{ 
Department of Physics, Tohoku University \\
Sendai 980-8578, Japan 
\\ {\rm E-mail: yamada@tuhep.phys.tohoku.ac.jp}}

\abstract{
We discuss some aspects of the radiative corrections in 
the phenomenology of the minimal SUSY standard model, by reviewing 
two recent studies. 
(1) The full one-loop corrections to the Higgs boson decays 
into charginos are presented, 
with emphasis on the renormalization of the chargino sector, including  
of their mixing matrices. (2) The two-loop 
$O(\alpha_s\tan\beta)$ corrections to the $b\to s\gamma$ decay 
in models with large $\tan\beta$, mainly those 
to the charged Higgs boson contributions, are discussed. 
Exact two-loop result is compared to an approximation 
used in previous studies. 
}

\normalsize\baselineskip=15pt

\section{Introduction}

There are many cases where the radiative corrections 
become important in the phenomenology of the 
minimal supersymmetric (SUSY) standard model (MSSM) \cite{mssm}.

(1) Of course, the radiative corrections become large when 
they are enhanced by large coupling constants and/or 
large logarithms. For example, QCD corrections to the processes 
involving quarks, gluon, and their superpartners, 
are indispensable in the study of the SUSY particles at 
hadron colliders. 

(2) Corrections to the observables which may be precisely 
measured in present or future experiments are also important. 
For example, electroweak precision measurements have provided 
a powerful tool to impose constraints on the SUSY particles. 
Also, the masses and couplings of several lighter SUSY particles 
are expected to be precisely measured at future linear 
colliders~\cite{LCexp}. 

(3) Radiative corrections may generate couplings 
which are strongly suppressed or even forbidden at 
lower levels of perturbation. 
As is well-known, the flavor-changing neutral current (FCNC) is 
forbidden at the tree-level of the standard model and 
sensitive to various types of new physics, including the SUSY particles. 
An example specific to the MSSM is the self-couplings of the 
Higgs bosons. The SUSY relation between the self-couplings and 
the electroweak gauge couplings is violated by loop corrections, 
resulting significant increase of the 
mass of the lightest Higgs boson $h^0$ \cite{mhiggs} beyond the 
theoretical upper limit at the tree-level. 

In this talk, we review two interesting recent studies of the radiative 
corrections in the MSSM phenomenology. 
In section 2, as a case of the class (2) listed above, 
the full one-loop corrections to the decays of heavier Higgs bosons 
into charginos are discussed, following Ref.~\cite{EMY}. The role of the 
renormalization of the chargino sector, including their mixing 
matrices, is explained in detail. 
In section 3, as a case of the class (3), 
two-loop $O(\alpha_s \tan\beta)$ SUSY QCD corrections to 
the ($b\to s\gamma$, $b\to sg$) decays in models with 
large $\tan\beta\equiv \langle H_U\rangle/\langle H_D \rangle$, 
especially the corrections to the charged Higgs boson 
contribution~\cite{BGY}, is discussed. 
Validity of the approximated calculation of the two-loop integrals, 
used in previous studies, is examined by comparison with 
the exact two-loop calculation. 

\section{One-loop Correction to the Chargino-Higgs boson Couplings}

Most of new particles in the MSSM, such as the SUSY particles and 
Higgs bosons, are mixtures of several gauge eigenstates \cite{mssm,GH}. 
Mixings of particles therefore play a crucial role in 
phenomenological studies of these particles. 

As an example, the charged SU(2) gauginos $\widetilde{W}^{\pm}_L$ and 
higgsinos $(\widetilde{H}_D^-, \widetilde{H}_U^+)_L$ mixes with each other 
to form two mass eigenstates $\tilde{\chi}^{\pm}_i(i=1,2)$, 
charginos, as 
\begin{equation}
\tilde{\chi}_{iL}^+ = V_{i\alpha}\left( \begin{array}{c} 
\widetilde{W}^+_L \\ \widetilde{H}_{UL}^+ \end{array} \right)_{\alpha}, \;\;\; 
\tilde{\chi}_{iL}^- = U_{i\alpha}\left( \begin{array}{c} 
\widetilde{W}^-_L \\ \widetilde{H}_{DL}^- \end{array} \right)_{\alpha} 
\;\;\; (i=1,2).
\end{equation}
At the tree-level, the mixing matrices $(V,U)$ are determined to 
diagonalize the mass matrix 
\begin{equation}
X = \left( \begin{array}{cc} M & \sqrt{2}m_W\sin\beta \\ 
\sqrt{2}m_W\cos\beta & \mu \end{array} \right) = 
U^T \left( \begin{array}{cc} m_{\tilde{\chi}^+_1} & 0 \\
                0 & m_{\tilde{\chi}^+_2} \end{array} \right) V . 
\label{eq:chi+mat}
\end{equation}
$M$ and $\mu$ are the mass parameters of the SU(2) gaugino and 
higgsinos, respectively. 
Couplings of the charginos are generally dependent on $(V,U)$. 

In future colliders, the masses and interactions of the charginos 
are expected to be measured precisely \cite{LCexp,LCs,charginomeasure}. 
It is therefore very interesting to study the radiative corrections to 
chargino interactions. In calculating the radiative corrections, 
we need to renormalize the chargino parameters, including the 
mixing matrices $(V,U)$. 
The renormalization of the chargino sector has been studied 
for different processes, such as 
$e^+e^-\to\tilde{\chi}^+\tilde{\chi}^-$ \cite{eechch,eechch2,eechch3}, 
$\tilde{f}\to f'\tilde{\chi}^{\pm}(f=q,l)$ \cite{sqch}, 
$H^+\to\tilde{\chi}^+\tilde{\chi}^0$ \cite{hpchne}, 
and 
$\tilde{\chi}^+\to\tilde{\chi}^0W^+$ \cite{chtow}. 

In this talk, we consider the decays of the heavier Higgs bosons 
($H^0$, $A^0$) into chargino pair, 
\begin{equation}
(H^0, A^0) \to \tilde{\chi}^+_i  + \tilde{\chi}^-_j \, ,
\label{eq:Hk0cha}
\end{equation}
with $i,j=(1,2)$. If $\tan\beta$ is not much larger than one, 
the decays (\ref{eq:Hk0cha}) may have non-negligible branching 
ratios \cite{tree1,tree2,tree3}.
These decays are also interesting because 
they are very sensitive to the mixings of charginos. 
Detailed studies of these decays, including radiative 
corrections, would therefore 
provide useful information about the chargino sector, 
complementary to the pair production processes 
$e^+e^-\to\tilde{\chi}^+_i\tilde{\chi}^-_j$
~\cite{chaproduction,LCs,charginomeasure}.

The tree-level widths of the decays (\ref{eq:Hk0cha}) are 
($H^0_{\{1,2,3\}}\equiv\{h^0, H^0, A^0\}$) 
\begin{eqnarray}
&& \hspace{-5mm} \Gamma^{\rm tree}(H_k^0 \to 
\tilde{\chi}^+_i \tilde{\chi}^-_j) =
\frac{g^2}{16 \pi\, m^{3}_{H_{k}^0} }\,
\kappa(m_{H_{k}^0}^2,m^{2}_{\tilde{\chi}^+_i},m^{2}_{\tilde{\chi}^+_j}) \,
\nonumber \\ 
&& \hspace{5mm}  \times
\left[ 
\left( m^{2}_{H_{k}^0} - m^{2}_{\tilde{\chi}^+_i} - 
m^{2}_{\tilde{\chi}^+_j} \right) 
(F_{ijk}^2 + F_{jik}^2 ) \, 
 - 4 \eta_k m_{\tilde{\chi}^+_i} m_{\tilde{\chi}^+_j} 
F_{ijk} F_{jik}  \right] \, ,
\end{eqnarray}
Here $\kappa(x,y,z)\equiv ((x-y-z)^2-4yz)^{1/2}$ and 
$\eta_k$ is the CP eigenvalue of $H_k^0$ ($\eta_{1,2}=1, \eta_3=-1$). 
Here we assume that the contributions of CP violation and 
generation mixings of the quarks and squarks are negligible. 
The tree-level couplings $gF_{ijk}$ of the Higgs bosons and 
charginos $H_k^0 \overline{\tilde{\chi}^+_{iL}} \tilde{\chi}_{jR}^+$ come 
from the gaugino-higgsino-Higgs boson couplings and 
take the forms \cite{GH}
\begin{eqnarray}
gF_{ijk} &=& \frac{g}{\sqrt{2}} 
\left(  e_k\, V_{i1}U_{j2} - d_k\, V_{i2}U_{j1} \right) \\ 
e_k&=&
 \Big(-\sin\alpha,\,\hphantom{-}\cos\alpha, 
\,-\sin\beta,\,\hphantom{-}\cos\beta \Big)_k\, , 
 \nonumber\\
 d_k&=&
 \Big(-\cos\alpha,\,-\sin\alpha,\,
\hphantom{-}\cos\beta,\,\hphantom{-}\sin\beta \Big)_k\, . 
 \label{eq:Fchtree}
\end{eqnarray}
Here $\alpha$ is the mixing angle for $(h^0, H^0)$. 
The Nambu-Goldstone mode $H^0_4\equiv G^0$ is included here 
for later convenience. 

The one-loop correction to the coupling $gF_{ijk}$ is 
expressed as 
\begin{equation}
gF^{\rm corr.}_{ijk} = 
gF_{ijk}+ \delta (gF^{(v)}_{ijk}) + g \delta F^{(w)}_{ijk} +
 \delta (gF^{(c)}_{ijk}) \, , 
\label{eq:Fren}
\end{equation}
where $\delta (gF^{(v)}_{ijk})$, $g\delta F^{(w)}_{ijk}$, and 
$\delta (gF^{(c)}_{ijk})$ are 
the proper vertex correction, the wave function correction to 
the external particles, and the 
counterterm by the renormalization of the 
parameters $(g,V,U,\alpha,\beta)$ in the tree-level coupling 
(\ref{eq:Fchtree}), respectively. 
The corrections from quarks and squarks in 
the third generation were calculated in Ref.~\cite{Zhang}. 
Here we present the full one-loop corrections 
shown in Ref.~\cite{EMY}, and show some numerical results for 
the $(A^0,H^0)\to\tilde{\chi}^+_1\tilde{\chi}^-_1$ decays. 

We discuss the wave function corrections $\delta F^{(w)}_{ijk}$ 
in detail. They are expressed as 
\begin{equation}
 \delta{F}_{ijk}^{(w)}\,=\, \frac{\,1}{\,2}\,\bigg[\,
\delta{Z}^{H^0}_{lk}{F}_{ijl}+ \delta{Z}_{i'i}^{+L}{F}_{i'jk}
+\delta{Z}_{j'j}^{+R}{F}_{ij'k}\bigg]
 \, .
\label{eq:dFw}
\end{equation}
$\delta{Z}^{+L}$ and $\delta{Z}^{+R}$ are corrections for the 
charginos, while $\delta{Z}^{H^0}_{lk}$ is for the Higgs bosons with 
$l=(1,2)$ for $k=(1,2)$ and $l=(3,4)$ for $k=3$. 
They are given in terms of the self-energies of 
the relevant particles. Explicit form of 
$\delta{Z}_{j'j}^{+L}$, wave function correction to the 
left-handed chargino $\tilde{\chi}^+_{jL}$, is given by 
\begin{eqnarray}
\lefteqn{\delta{Z}^{+L}_{ii}  = } && \nonumber\\
&& \hspace*{-5mm} - {\rm Re}\,
 \bigg\{\Pi^{\tilde{\chi} L}_{ii}(m_i^2)+ m_i\,
 \Big[m_i\dot{\Pi}^{\tilde{\chi} L}_{ii}(m_i^2)
 +m_i\dot{\Pi}^{\tilde{\chi} R}_{ii}(m_i^2)
 +2\dot{\Pi}^{\tilde{\chi} S,L}_{ii}(m_i^2) 
 \Big]\bigg\}\, ,
 \label{eq:dZchpp}
 \\
\lefteqn{  \delta{Z}^{+L}_{pi} = } && \nonumber \\
&& \hspace*{-5mm}\frac{2}{m_p^2-m_i^2}\;
 {\rm Re}\left\{
 m_i^2 \Pi^{\tilde{\chi} L}_{pi}(m_i^2) 
 + m_i m_p \Pi^{\tilde{\chi} R}_{pi}(m_i^2) + 
 m_p \Pi^{\tilde{\chi}\,S,L}_{pi}(m_i^2) 
 + m_i \Pi^{\tilde{\chi}\,S,R}_{pi}(m_i^2) \right\} 
\, ,
 \label{eq:dZchps}
\end{eqnarray}
where $p\neq i$ and 
\begin{equation}
\Pi^{\tilde{\chi}}_{ij}(p)=\Pi^{\tilde{\chi} L}_{ij}(p^2)
{p \hspace{-1.8mm} \slash} P_L 
+\Pi^{\tilde{\chi} R}_{ij}(p^2)
{p \hspace{-1.8mm} \slash} P_R 
+\Pi^{\tilde{\chi}\,S,L}_{ij}(p^2)P_L+\Pi^{\tilde{\chi}\,S,R}_{ij}(p^2)P_R \, ,
\label{eq:charginoselfe} 
\end{equation}
are the self-energies of the charginos $\tilde{\chi}^+$. 
$\delta Z^{+R}$ for the right-handed chargino $\tilde{\chi}^+_R$ 
is obtained from Eqs.~(\ref{eq:dZchpp}, \ref{eq:dZchps})
by the exchange $L\leftrightarrow R$. 
We used the CP symmetry relation ${\rm Re}\Pi^{\tilde{\chi} S,L}_{ii}
={\rm Re}\Pi^{\tilde{\chi} S,R}_{ii}$ 
in Eq.~(\ref{eq:dZchpp}). 
The corrections $\delta{Z}^{H^0}$ are
\begin{eqnarray}
 \label{eq:dZHkk}
 \delta{Z}^{H^0}_{kk}&=&
  -\;{\rm Re}\,\dot{\Pi}^{H^0}_{kk}(m_{H_k^0}^2)\, , 
\hspace{30mm} k=1,2,3, \\ 
\label{eq:dZHlk}
\delta{Z}^{H^0}_{ab}&=& \frac{2}{m^2_{H^0_a}-m^2_{H^0_b}}\, {\rm
Re}\, \Pi^{H^0}_{ab}(m^2_{H^0_b}) \, , 
\hspace{10mm} a,b=(1,2), \;  a\neq b \\
\label{eq:dZAG}
\delta{Z}^{H^0}_{43}&=& -\frac{2}{m_{A^0}^2}\, {\rm
Re}\, \Pi^{H^0}_{43}(m_{A^0}^2) \, . 
\end{eqnarray}
The Higgs boson self-energies $\Pi^{H^0}(k^2)$ in 
Eqs. (\ref{eq:dZHkk}, \ref{eq:dZHlk}, \ref{eq:dZAG}) include 
momentum-independent contributions from the 
tadpole shifts~\cite{oshiggs} and leading 
higher-order corrections. 
The latter contribution is numerically relevant for  
$h^0$ and $H^0$. 
Note that $\delta{Z}^{H^0}_{43}$ in Eq.~(\ref{eq:dZAG}) 
includes both the $A^0-G^0$ and $A^0-Z^0$ mixing contributions. 

The off-diagonal part of the wave function correction 
$\delta Z_{ij}(i\neq j)$ is generated by the mixing between 
the tree-level mass eigenstates at the one-loop level, and 
closely related to the renormalization of the mixing matrices. 
To see this point, we focus on the contribution of 
$\delta{Z}_{i'i}^{+L}$ in Eq.~(\ref{eq:dFw}) and 
decompose $\delta{Z}^{+L}$ into hermitian and 
anti-hermitian parts, to obtain 
\begin{eqnarray}
\frac{1}{2} ( \delta{Z}_{ii}^{+L}{F}_{ijk} + \delta{Z}_{pi}^{+L}{F}_{pjk})
&=& 
\frac{1}{2}\delta{Z}_{ii}^{+L}{F}_{ijk} + 
\frac{1}{4}[ \delta{Z}_{pi}^{+L} + (\delta{Z}_{ip}^{+L})^* ] {F}_{pjk}
\nonumber \\
&& + \frac{1}{4}[ \delta{Z}_{pi}^{+L} - (\delta{Z}_{ip}^{+L})^* ] {F}_{pjk} .
\label{cancellation}
\end{eqnarray}
The ultraviolet (UV) divergence of the hermitian part in the 
first line is cancelled by that of the vertex correction 
$\delta F_{11k}^{(v)}$ and the counterterm $\delta g$. 
On the other hand, the divergence of the anti-hermitian part 
in the second line is cancelled by the counterterm $\delta V$ for 
the mixing matrix $V$ of $\tilde{\chi}^+_L$, giving 
\begin{equation}
\delta F^{(c)}_{ijk}(\delta V) = 
(\delta V\cdot V^{\dagger})_{ip} F_{pjk} .
\label{cancellation2}
\end{equation}
The matrix $\delta V\cdot V^{\dagger}$ should be anti-hermitian 
for the unitarity of $V$ and $V^{\rm bare}\equiv V+\delta V$. 
Similarly, the UV divergences of the anti-hermitian 
parts of $\delta Z^{+R}$ and $\delta Z^{H^0}$ are cancelled by 
renormalization of $U$ and 
$\alpha$ (for $H^0$, $h^0$) or $\beta$ (for $A^0$), respectively. 
This relation between the UV divergence of the anti-hermitian part 
of the wave function corrections $\delta Z$ and 
the renormalization of the corresponding mixing 
matrix holds for general cases \cite{earlier,gaugedepavoid,later}. 

To fix the chargino sector, we have to specify two input parameters 
corresponding to two parameters ($M$, $\mu$) in the mass 
matrix (\ref{eq:chi+mat}), in addition to $\tan\beta$ which is 
determined by the Higgs boson sector. 
The pole masses $m_{\tilde{\chi}^+_i}$ and 
renormalized mixing matrices $(V,U)^{(ren)}$ are 
then given as functions of these input parameters. 
We also need to fix a definition of the renormalized 
mixing matrices, or the UV finite parts of 
the counterterms ($\delta V$, $\delta U$). 
In previous studies of the corrections to chargino interactions, 
several schemes has been proposed for the 
renormalization of the charginos, as listed below: 

(A) We may just use the running mass parameters $(M,\mu)$ in the 
$\overline{\rm DR}$ scheme at a scale $Q$ as inputs, as in 
Ref.~\cite{eechch}. 
Renormalized $(V,U)$ are fixed to diagonalize 
the tree-level mass matrix. The pole masses $m_{\tilde{\chi}^+_i}$ are 
shifted from their $Q$-dependent tree-level values. 
The effect of this mass shift has to be 
taken into account for a proper treatment of the radiative corrections 
to chargino processes. 

(B) On the other hand, one may fix the chargino sector by 
specifying the pole masses of two charginos $m_{\tilde{\chi}^+_i}(i=1,2)$, 
as in Ref.~\cite{chmasscorr3}. 
Renormalized $(M,\mu)$ are then defined as tree-level functions of 
the pole masses. Again, renormalized $(V,U)$ diagonalize the 
tree-level mass matrix. In this scheme, the pole masses of charginos are 
identical to their tree-level values by definition. 
However, one should note that the shift of the masses is unavoidable 
when the neutralinos $\tilde{\chi}^0_i(i=1-4)$ appear in the analysis, 
since there are only three free parameters $(M,\mu,M')$ to 
describe two charginos and four neutralinos. 

(C) Alternatively, we may start from the ``on-shell mixing matrices'' 
$(V,U)^{\rm OS}$ \cite{earlier,chmasscorr1}, 
defined such that their counterterms 
completely cancel the anti-hermitian part of the 
corresponding wave function corrections. For example, the 
second line of Eq.~(\ref{cancellation}) is 
dropped by adding Eq.~(\ref{cancellation2}) with on-shell $\delta V$. 
The renormalized $(M,\mu)$ are then given as diagonal elements of the 
``on-shell mass matrix'' $X^{OS}$ of the charginos \cite{chmasscorr1} 
given as 
\begin{equation}
X^{\rm OS} = \left( \begin{array}{cc} M^{\rm OS} & X^{\rm OS}_{12} \\ 
X^{\rm OS}_{21} & \mu^{\rm OS} \end{array} \right) 
\equiv 
(U^{\rm OS})^T \left( \begin{array}{cc} m_{\tilde{\chi}^+_1} & 0 \\
                 0 & m_{\tilde{\chi}^+_2} \end{array} \right)_{\rm pole} 
V^{\rm OS} 
\end{equation}
The off-diagonal elements $(X_{12}, X_{21})^{\rm OS}$ include 
some information of the loop corrections to the mixings and, 
as a result, deviate from their tree-level 
values $\sqrt{2}m_W(\sin\beta,\cos\beta)$. 
In this scheme, however, 
both the masses $m_{\tilde{\chi}^+_i}$ and mixing matrices ($V$, $U$) 
are shifted from their tree-level values. 
Problem from the gauge parameter dependence of the 
on-shell mixing matrices \cite{gaugedep,gaugedepavoid,later} 
may be avoided by improving relevant self energies 
by the pinch technique \cite{gaugedepavoid,espinosa}. 

We conclude this section with several numerical results, 
adopted from Ref.~\cite{EMY}, 
for the decay widths of $(A^0,H^0)\to\tilde{\chi}^+_1+\tilde{\chi}^-_1$ 
in the renormalization scheme (C) shown above. 
Calculation was done by using the packages 
{\it FeynArts, FormCalc,} and {\it LoopTools} \cite{FeynArts}. 
We use the SPS1a parameter point \cite{SPS1a} as reference point: 
Chargino and neutralino sectors are specified by 
the on-shell parameters $M=197.6$~GeV, $\mu=353.1$~GeV, $M'=98$~GeV, 
and the on-shell parameters for Higgs boson sector, defined as 
Ref.~\cite{oshiggs}, are $\tan\beta=10$ and $m_{A^0}=393.6$~GeV. 
The SUSY-breaking sfermion-Higgs boson trilinear couplings 
$(A_t,A_b,A_\tau)=(-487,-766,-250)$~GeV are given in the 
$\overline{\rm DR}$ scheme at the parent particle. 
Other mass parameters for sfermions are 
$(M_{\tilde Q_{1,2}},M_{\tilde U_{1,2}},M_{\tilde D_{1,2}},
M_{\tilde L_{1,2}},M_{\tilde E_{1,2}}) = 
(558.9,540.5,538.5,197.9,137.8)$~GeV 
for the first and second generations and 
$(M_{\tilde Q_{3}},M_{\tilde U_{3}},$ $M_{\tilde D_{3}},
M_{\tilde L_{3}},M_{\tilde E_{3}}) = 
(512.2, 432.8, 536.5, 196.4, 134.8)$~GeV 
for the third generation. 
We used these values in the figures of this section, 
if not specified otherwise. 
Using HDECAY program \cite{HDECAY}, the tree-level branching ratios 
${\rm Br}(A^0\to\tilde{\chi}^+_1\tilde{\chi}^-_1)$ and 
${\rm Br}(H^0\to\tilde{\chi}^+_1\tilde{\chi}^-_1)$ 
at this point are estimated to be 21\% and 4\%, respectively, 
which are not negligible. 

\begin{figure}[htb]
\begin{center}
\includegraphics[width=7.5cm]{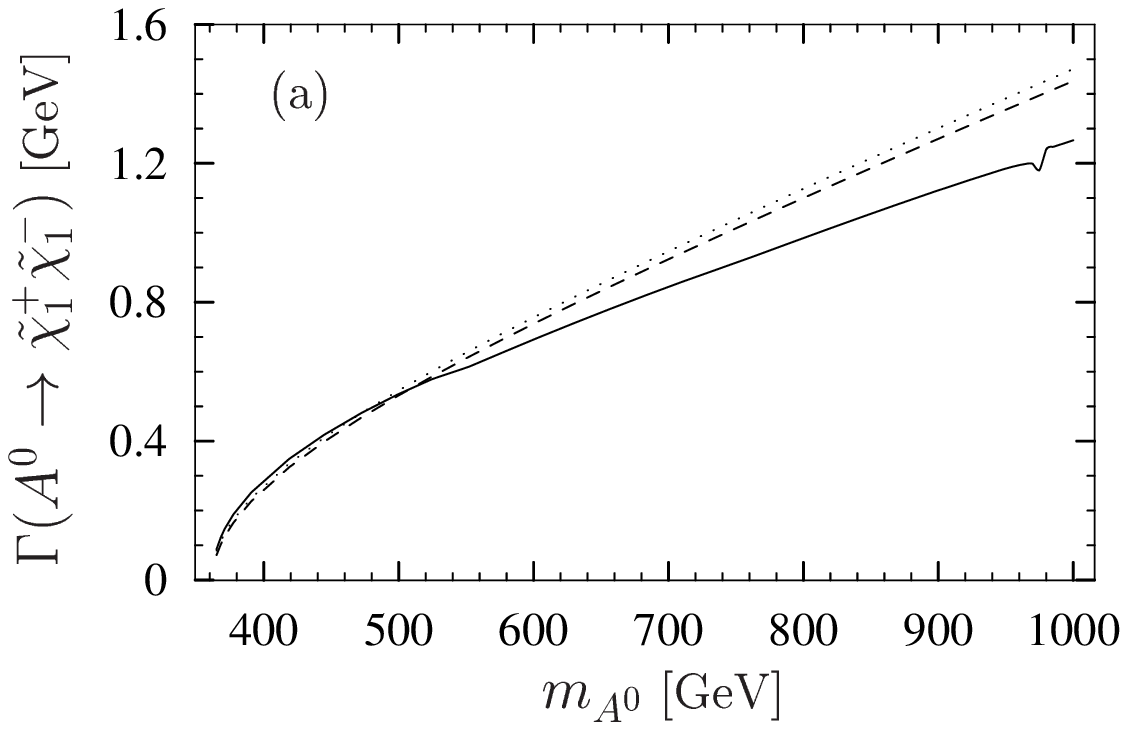}
\hspace{4mm}
\includegraphics[width=7.5cm]{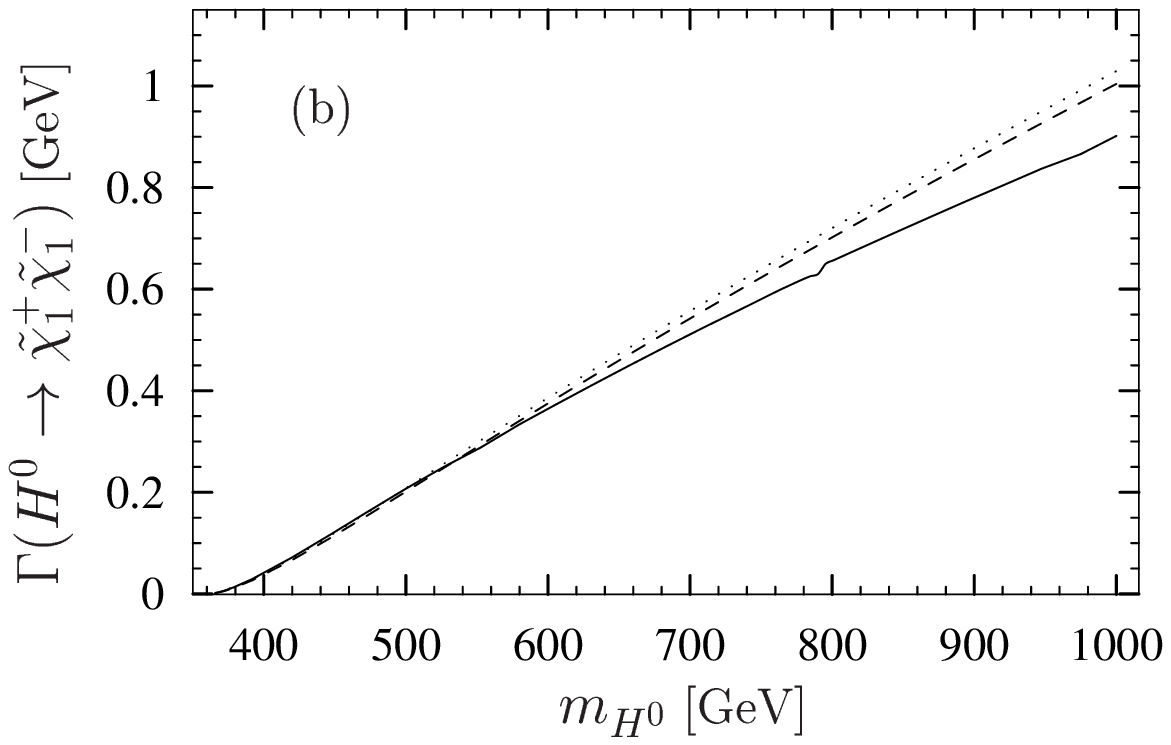}
\caption{%
Naive tree-level (dotted), tree-level (dashed), and 
one-loop corrected (solid) widths of 
$A^0\to\tilde{\chi}^+_1\tilde{\chi}^-_1$ (a) and 
$H^0\to\tilde{\chi}^+_1\tilde{\chi}^-_1$ (b) 
as functions of the parent particle. }
\label{figa1}
\end{center}
\end{figure}

In Fig.~\ref{figa1}, we show the decay widths of 
$A^0\to\tilde{\chi}^+_1\tilde{\chi}^-_1$ and 
$H^0\to\tilde{\chi}^+_1\tilde{\chi}^-_1$
as functions of the parent particle, 
and compare three definitions of the widths: 
the naive tree-level width $\Gamma^{\rm naive}$ with 
the tree-level $m_{\tilde{\chi}^+_i}$ and $(V,U)$, 
the tree-level width $\Gamma^{\rm tree}$ using 
the pole masses $m_{\tilde{\chi}^+_i}$ and $(V,U)^{OS}$, 
and the full one-loop corrected width $\Gamma^{\rm corr}$ which also 
includes real photon emission 
$(A^0,H^0)\to\tilde{\chi}^+_1\tilde{\chi}^-_1\gamma$ 
to cancel infrared divergence. 
We see that the full one-loop corrections amount up to $\sim -12$\%. 

\begin{figure}[htb]
\begin{center}
\includegraphics*[width=10cm]{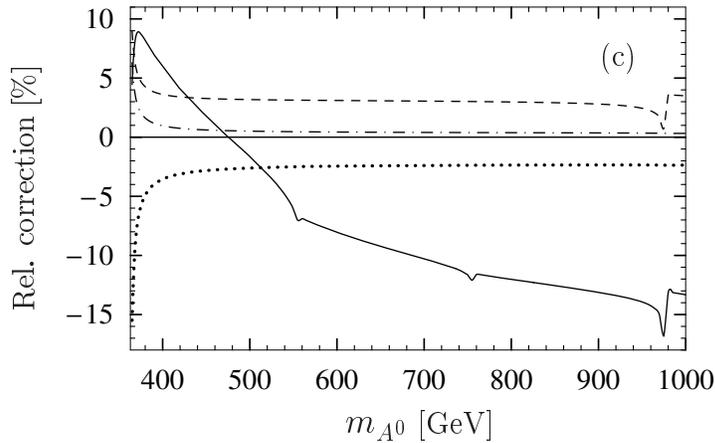}
\caption{%
Indivisual one-loop corrections to the decay width of 
$A^0\to\tilde{\chi}^+_1\tilde{\chi}^-_1$ relative to the 
naive tree-level width. Explanation of each line is seen in the text. 
}
\label{figa2}
\end{center}
\end{figure}

In Fig.~\ref{figa2} we compare the contributions from the (s)fermion 
loops~\cite{Zhang} (loops with quarks, leptons, and their superpartners) 
and the full one-loop contributions, relative to $\Gamma^{\rm naive}$, 
for Fig.~\ref{figa1}(a). 
Corrections to the chargino mass matrix are shown by 
the dash-dotted line for the (s)fermion loops while 
the dotted line is for the full correction. 
The solid (dashed) line shows the total correction 
including full ((s)fermion) one-loop contributions. 
Figure \ref{figa2} shows that the (s)fermion loop corrections and 
other corrections are of comparable order, both for the chargino 
mass matrix and for the conventional loop corrections (\ref{eq:Fren}). 

\begin{figure}[htb]
\begin{center}
\hspace*{2mm}
\includegraphics[width=73mm]{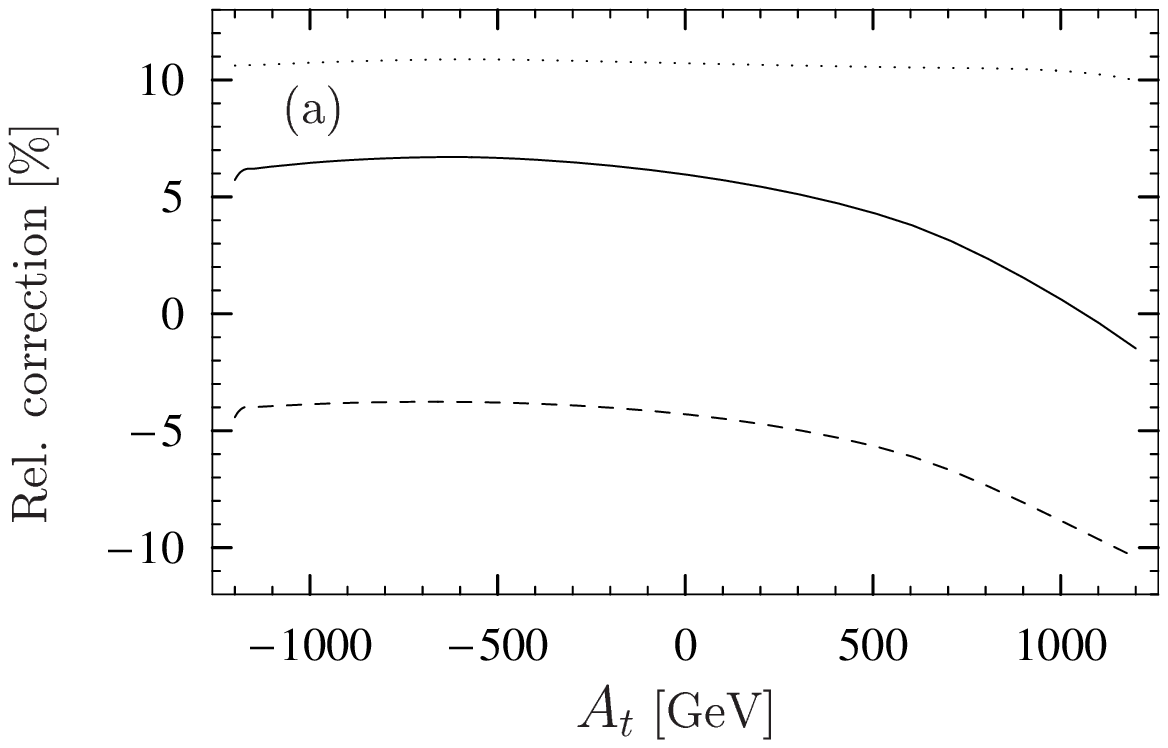}
\hspace{5mm}
\includegraphics[width=73mm]{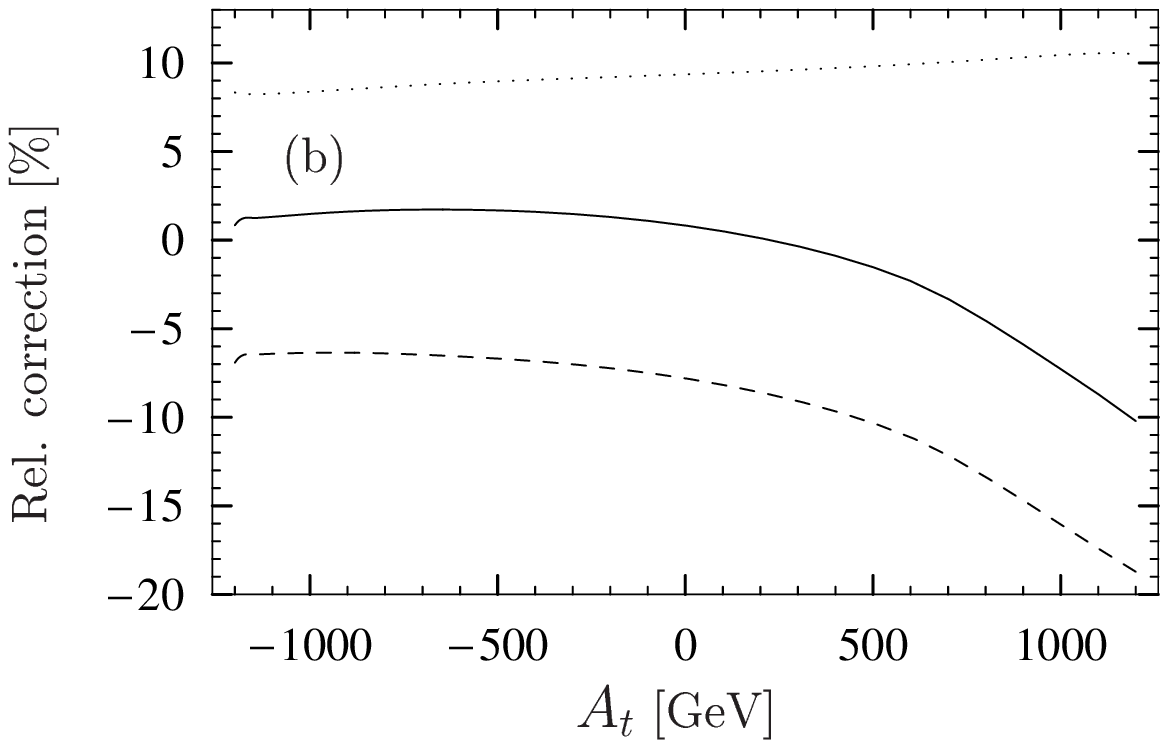}
\caption{ %
Relative corrections for the decays 
$A^0\to\tilde{\chi}^+_1+\tilde{\chi}^-_1$~(a) 
and $H^0\to\tilde{\chi}^+_1+\tilde{\chi}^-_1$~(b) 
as functions of $A_t=A_b=A_{\tau}$. 
The dashed lines, solid lines, and dotted lines 
denote $\Gamma^{\rm tree}/\Gamma^{\rm naive}-1$, 
$\Gamma^{\rm corr}/\Gamma^{\rm naive}-1$, and 
$\Gamma^{\rm corr}/\Gamma^{\rm tree}-1$, respectively. }
\label{fig:Adependence}
\end{center}
\end{figure}

Fig.~\ref{fig:Adependence} shows the corrections to the decay 
widths of $(A^0,H^0)\to\tilde{\chi}^+_1+\tilde{\chi}^-_1$ 
as a function of $A_t = A_b = A_\tau$, 
with the other parameters unchanged.
The dashed lines denote $\Gamma^{\rm tree}/\Gamma^{\rm naive}-1$ and 
show the effect of the chargino mass matrix correction. 
The solid lines show the total correction 
$\Gamma^{\rm corr}/\Gamma^{\rm naive}-1$. 
The dotted lines stand for $\Gamma^{\rm corr}/\Gamma^{\rm tree}-1$, 
the conventional loop correction in Eq.~(\ref{eq:Fren}). 
One sees that the $A_t$ dependence of the corrected widths 
mainly comes through the shifts of the masses and 
mixing matrices of the charginos from the tree-level values. 

\section{Two-loop $O(\alpha_s\tan\beta)$ Corrections 
to $b\to s\gamma$} 

There are many cases where two-loop and even higher order 
radiative corrections are necessary in the MSSM phenomenology. 
As a well-known example, the correction to the mass of the 
lightest Higgs boson $h^0$ is so large that the two-loop 
contribution~\cite{mhiggs2,mhiggs2v} is still larger than 
the expected error in future measurements. 

Here we consider the two-loop $O(\alpha_s\tan\beta)$ SUSY QCD corrections to 
the $b\to s\gamma$ and $b\to sg$ decays in models with large $\tan\beta$. 
These decays describe the inclusive decay 
width ${\rm Br}(\bar{B}\to X_s \gamma)$ very well ~\cite{NLOSM}, 
up to the nonperturbative hadronic corrections which are small 
and well under control. 

In the standard model, the decays $b\to(s\gamma, sg)$ 
occur through $W^{\pm}$ boson loops. These decays are 
important to prove possible new physics beyond the standard model since 
the new physics may contribute at the same level of perturbation as 
the standard model one. 

In the MSSM, these decays receive new 
contributions~\cite{BSGinSUSYproposal,BSGcontributions} 
from loops with the charged Higgs boson $H^\pm$, 
charginos $\tilde{\chi}^{\pm}$, gluino $\tilde{g}$, 
and neutralino. Their contributions are often comparable to 
or even larger than the $W^{\pm}$ loop, 
and sensitive to the masses and couplings 
of these new particles. The leading order 
QCD corrections to these new contributions 
have been calculated~\cite{gluino} for generic models. 
Higher-order QCD and SUSY QCD corrections have been evaluated 
for specific models~\cite{NLO-SUSY1,NLO-SUSY2a,NLO-SUSY2b}. 

Here we are interested in the $b\to(s\gamma,sg)$ decays in 
models with very large $\tan\beta$~\cite{bsglargeTB1,bsglargeTB2}. 
One important finding is that the SUSY QCD may induce 
${\cal O}(\alpha_s\tan\beta)$ corrections~\cite{NLO-SUSY2a,NLO-SUSY2b} 
to the contributions of the charged Higgs boson and of charginos. 
These two-loop corrections may be comparable to the leading 
one-loop contributions, as shown below,  and significantly affect the 
experimental constraints~\cite{bsglargeTB1,bsglargeTB2} on the new particles. 
In this talk, we mainly consider the corrections to the contribution of 
the charged Higgs boson $H^+$, following Ref.~\cite{BGY}. 

\begin{figure}[h] 
\vspace{0.3truecm}
\begin{center} 
\includegraphics[width= 6.5cm]{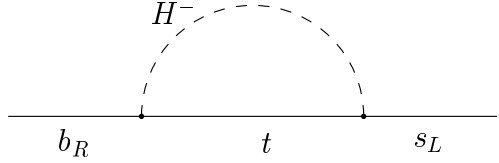}
\end{center} 
\caption[f1]{
$b\to s\gamma$ and $b\to sg$ decays by 
the one-loop $H^{\pm}$ exchange. 
The photon or gluon is to be attached at the $t$ or $H^-$ 
lines.}
\label{CH-0}
\end{figure}

At the one-loop level, dominant contribution by $H^{\pm}$ exchange 
comes from the diagram 
in Fig.~\ref{CH-0} with initial $b_R$. 
Couplings of $H^{\pm}$ to quarks are derived from the tree-level 
lagrangian 
\begin{equation}
{\cal L}_{\rm int} = -h_b \bar{b}_R q_L H_D  
-h_t \bar{t}_R q_L H_U +({\rm h.c.}), 
\label{treelag}
\end{equation}
where only quarks in the third generation $(t,b)$ are included 
for simplicity. 
At the tree-level, the couplings of $H_U$ to $b_R$ and of $H_D$ to 
$t_R$ are forbidden by SUSY and the Peccei-Quinn symmetry 
under $(q_L,t_R,H_U)\to(q_L,t_R,H_U)$, $(b_R,H_D)\to (-b_R,-H_D)$. 
However, squark-gluino loops with breakings of both symmetries 
may induce effective couplings \cite{dmb,carenaH0,Ambrosio}
\begin{equation}
\Delta {\cal L}_{\rm eff. int} = 
-h_b\Delta_b \bar{b}_Rq_L H_U -h_t\Delta_t \bar{t}_Rq_L H_D
+({\rm h.c.}) .
\label{efflag}
\end{equation} 
$\Delta_q(q=b,t)$ are one-loop functions of 
$O(\alpha_s \mu m_{\tilde{g}}/M_{\rm SUSY}^2)$, where squarks and gluino 
masses are around the scale $M_{\rm SUSY}$. There are also 
$O(h_t^2)$ contributions to Eq.~(\ref{efflag}) from squark-higgsino 
loops. Note that $\Delta_q$ do not 
decouple in large $M_{\rm SUSY}$ limit \cite{dmb}. 

Although $|\Delta_q|$ themselves are sufficiently smaller than 
unity, their contributions to the $H^+$ couplings are 
enhanced by $\tan\beta$ relative to the tree-level, 
as shown below, and may give large corrections 
in large $\tan\beta$ models. 

(i) Correction from counterterm to $m_b$ \cite{dmb}: 
The QCD running mass $m_b({\rm SM})$ within the standard model 
is given by Eqs.~(\ref{treelag}, \ref{efflag}) as 
\begin{eqnarray}
m_b({\rm SM}) &=& 
\frac{h_b\bar{v}}{\sqrt{2}}\cos\beta[1+\Delta_b \tan\beta ] \\
&=& m_b({\rm MSSM})+\delta m_b .
\label{eq:dmb}
\end{eqnarray}
The squark-gluino correction $\delta m_b$ 
lift tree-level suppression of $m_b$ by $\cos\beta$ and may become 
comparable to the tree-level contribution. 
As a result, the $H^+\bar{t}_Lb_R$ coupling $y_b$ may significantly 
deviate from the tree-level as 
\begin{eqnarray}
y_b(H^+\bar{t}_Lb_R)({\rm eff}) &=& 
V_{tb}h_b\sin\beta(1-\Delta_b \cot\beta) \nonumber \\
&\rightarrow & 
V_{tb}\frac{\sqrt{2}m_b({\rm SM})}{\bar{v}}\tan\beta
\frac{1}{1+\Delta_b \tan\beta } .
\end{eqnarray}
The large correction $\Delta_b\tan\beta$ is originated from that 
the $H^+\bar{t}_Lb_R$ coupling receives very small contribution from 
Eq.~(\ref{efflag}) because of 
$H^+=\sin\beta H^+_D+\cos\beta H^+_U\sim H_D^+$. 
Similarly, $\delta m_b$ in Eq.~(\ref{eq:dmb}) also induce 
$O(\alpha_s\tan\beta)$ corrections 
to the couplings of $b_R$ to heavier Higgs bosons 
($H^0$, $A^0$) and to the higgsino $\widetilde{H}_D$. 

(ii) Correction to the $H^-\bar{b}_Lt_R$ coupling $y_t$ 
comes from $\Delta_t$ through the proper vertex correction as 
\cite{carenaH0,Ambrosio,Burashuge} 
\begin{eqnarray}
y_t(H^+\bar{b}_Lt_R)({\rm eff}) &=& 
V_{tb}^*h_t\cos\beta(1-\Delta_t \tan\beta) \\
&\to & 
V_{tb}^*\frac{\sqrt{2}m_t}{\bar{v}}\cot\beta
(1-\Delta_t \tan\beta) .
\end{eqnarray}

In general, Eq.~(\ref{efflag}) has mixing terms between quarks 
in different generations, which are induced by the squark-higgsino 
loops and squark-gluino loops with squark generation 
mixings \cite{dmb,Burashuge}. 
These mixing terms may generate $\tan\beta$-enhanced 
corrections to the CKM matrix $V$ 
and flavor-changing couplings of ($H^0$, $A^0$). The latter couplings 
induce the decays $B_s\to \mu^+\mu^-$ \cite{btomumu,Burashuge} 
and $(H^0,A^0)\to b\bar{s}$ \cite{htobs}. 

\begin{figure}[t] 
\begin{center} 
\includegraphics[width= 6.0cm]{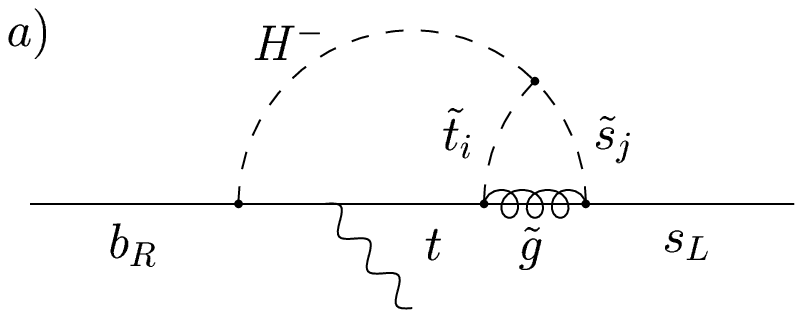}
\hspace{5mm}
\includegraphics[width= 6.0cm]{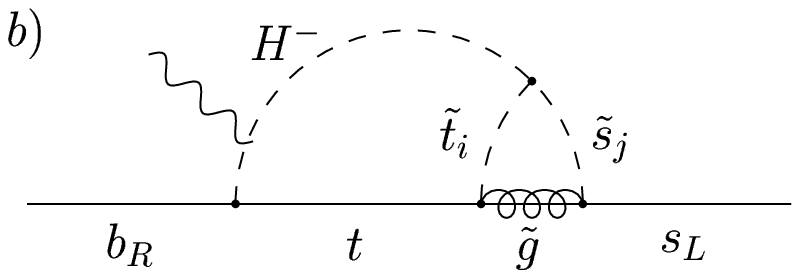}
\end{center} 
\begin{center} 
\includegraphics[width= 6.0cm]{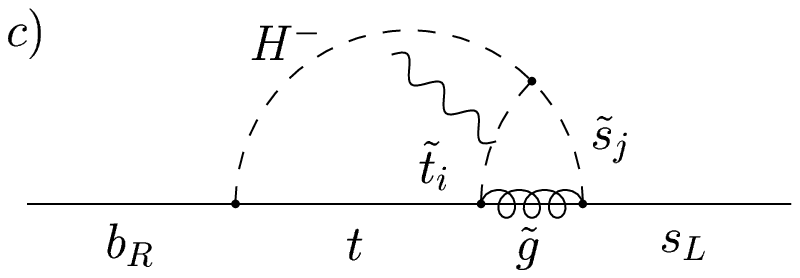}
\hspace{5mm}
\includegraphics[width= 6.0cm]{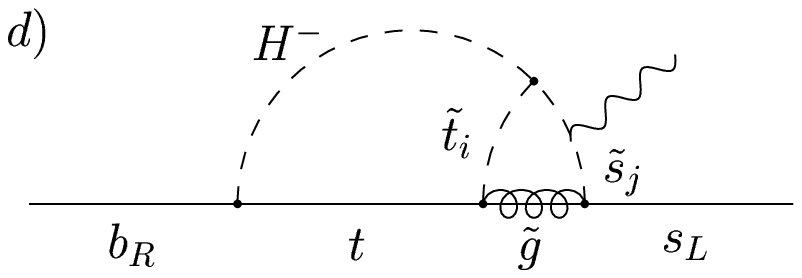}
\hspace{5mm}
\includegraphics[width= 6.0cm]{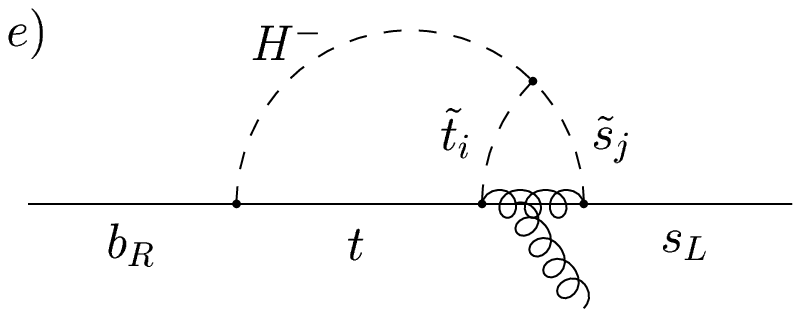}
\end{center} 
\vspace{-0.2truecm}
\caption{ $H^\pm$ mediated diagrams contributing 
 at order ${\cal O}(\alpha_s \tan \beta)$ to the decays 
 $b\to s\gamma$ and $b\to s g$. The photon must be replaced by a 
 gluon and vice versa, whenever possible.} 
\label{CHexch2loops}
\end{figure}

Two-loop ${\cal O}(\alpha_s\tan\beta)$ corrections to the 
$b\to (s\gamma, sg)$ decays has been calculated in Refs. 
\cite{NLO-SUSY1,NLO-SUSY2a,NLO-SUSY2b,Burashuge}. 
Here we discuss the $H^{\pm}$ contributions 
to the Wilson coefficients $C_i(\mu)(i=7,8)$, 
defined in the effective Hamiltonian 
\begin{equation}
H_{\rm eff} \supset
 -\frac{4G_F}{\sqrt{2}}V^*_{ts}V_{tb}
 \left( C_7(\mu) {\cal O}_7(\mu) + C_8(\mu) {\cal O}_8(\mu) 
 \right) \, ,
\end{equation}
with 
\begin{equation}
{\cal O}_7(\mu) = 
 \frac{e}{16\pi^2} m_b(\mu)\bar{s}_L\sigma^{\mu\nu} b_R F_{\mu\nu}\,, 
\hspace{0.8truecm}
{\cal O}_8(\mu) = 
 \frac{g_s}{16\pi^2}m_b(\mu)\bar{s}_L\sigma^{\mu\nu}T^a b_RG^a_{\mu\nu}\,. 
\label{O7and8}
\end{equation}
The $H^{\pm}$ contributions $C_{i,H}(i=7,8)$ to $O(\alpha_s\tan\beta)$ 
at the scale $\mu_W=m_W$ are expressed as 
\begin{equation}
C_{i,H}(\mu_W) = \frac{1}{1+\Delta_{b_R,b} \tan\beta}
\left[ C_{i,H}^0(\mu_W) + \Delta C_{i,H}^1(\mu_W) \right] \, . 
\label{defWC}
\end{equation}
Here $C_{i,H}^0(\mu_W)$ and $\Delta C_{i,H}^1(\mu_W)$
are the contributions of the one-loop diagram 
and the two-loop diagrams in Fig.~\ref{CHexch2loops}, respectively. 
The overall factor $1/(1+\Delta_{b_R,b} \tan\beta)$ represents 
the correction from $\delta m_b$. The one-loop integral 
$\Delta_{b_R,b}$ improves $\Delta_b$ in 
Eq.~(\ref{efflag}) by inclusion of the SU(2)$\times$U(1) 
breaking for the masses and couplings of 
squarks \cite{NLO-SUSY2a,Burashuge,BGY}. 

In previous studies~\cite{NLO-SUSY1,NLO-SUSY2a,NLO-SUSY2b}, 
the $O(\alpha_s \tan\beta)$ SUSY QCD corrections were evaluated in terms of 
an effective two-Higgs-doublet lagrangian, in which squarks and gluino 
are integrated out. This approach is called the 
``nondecoupling approximation'' in Ref.~\cite{BGY} since it preserves 
all $O(M_{\rm SUSY}^0)$ contributions of the original two-loop 
integrals. 
For the corrections (i) from $\delta m_b$, this approximation allows 
us to resum higher-order $O((\alpha_s\tan\beta)^n)$ terms 
(Carena et al. in Ref.~\cite{carenaH0}), by putting 
$\Delta_{b_R,b}$ in the denominator as in Eq.~(\ref{defWC}). 
In contrast, for the proper vertex corrections (ii) 
to the $H^-\bar{s}_Lt$ coupling in Fig.~\ref{CHexch2loops}, 
the nondecoupling approximation retains only diagrams (a) and (b), and the 
squark-gluino subloops are evaluated at vanishing external momenta. 
The $O(\alpha_s\tan\beta)$ result (\ref{defWC}) is then 
approximated by a rather simple form 
\begin{equation}
C_{i,H}(\mu_W)|_{\rm nondec} = 
\frac{1-\Delta_{t_R,s}\tan\beta }{1+\Delta_{b_R,b} \tan\beta}
C_{i,H}^0(\mu_W) \, . 
\label{defWCnondec}
\end{equation}
The one-loop integral $\Delta_{t_R,s}$, defined in Ref.~\cite{BGY}, 
corresponds to $\Delta_t$ in Eq.~(\ref{efflag}) applied for the 
$H^-\bar{s}_Lt_R$ coupling, including the SU(2)$\times$U(1) 
breaking effect \cite{Burashuge}. 

However, the momentum dependence of the squark-gluino subloops in 
Fig.~\ref{CHexch2loops}a,b), as well as the 
diagrams in Fig.~\ref{CHexch2loops}c-e) 
ignored in the nondecoupling approximation, are expected to give 
$O((m_{\rm weak}^2, m_{H^{\pm}}^2)/M_{\rm SUSY}^2)$ contributions, 
where $m_{\rm weak}\sim(m_W,m_t)$, and, therefore, cause 
significant deviation of the exact two-loop result from 
the nondecoupling approximation 
when $M_{\rm SUSY}$ is not much larger than 
$m_{\rm weak}$ and/or $m_{H^{\pm}}$. 
It is important to examine, in such cases, 
how large the deviation is and how far the nondecoupling 
approximation may be applied beyond the restriction 
$(m_{\rm weak}^2, m_{H^{\pm}}^2)\ll M_{\rm SUSY}^2$. 

\begin{figure}[htb]
\begin{center} 
\includegraphics*[width= 7cm]{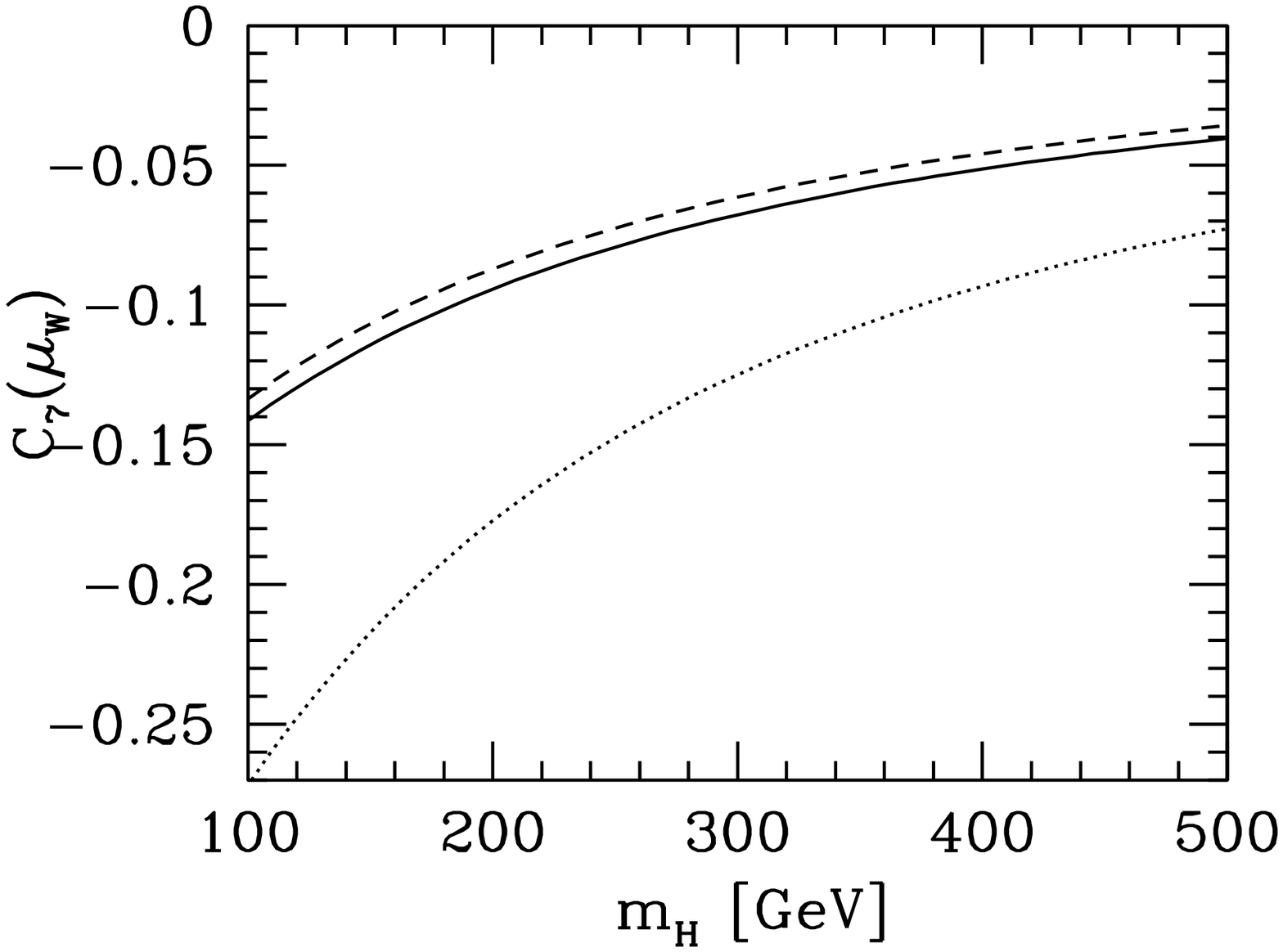}
\includegraphics*[width= 7cm]{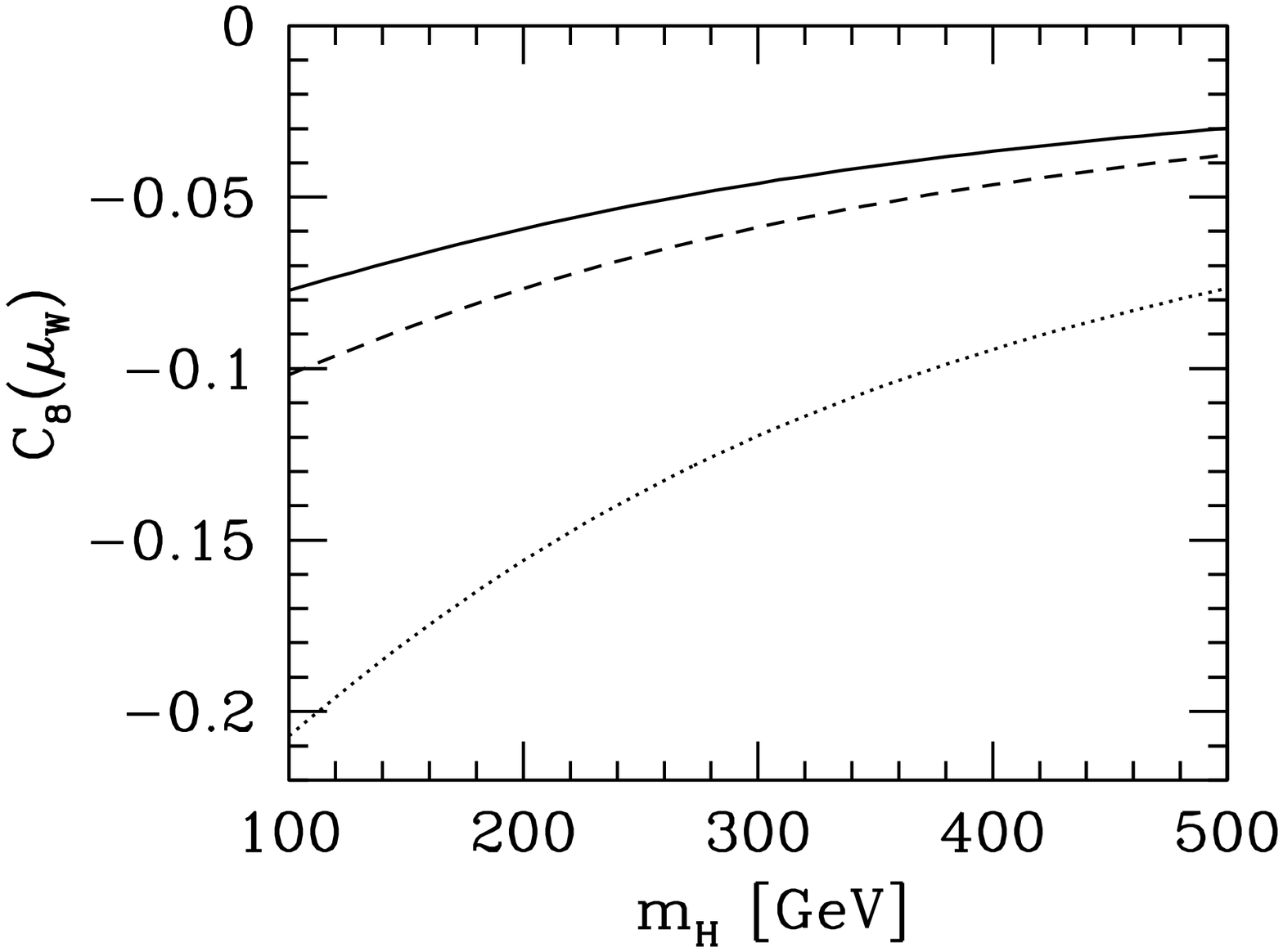} 
\end{center} 
\caption{%
$C_{7,H}(\mu_W)$ and $C_{8,H}(\mu_W)$ as functions of $m_H$. 
The dotted, dashed, and solid lines show the one-loop result, 
nondecoupling approximation, and exact two-loop result, respectively. 
Parameters for the SUSY particles are shown in the text. 
}
\label{figb1}
\end{figure}

\begin{figure}[htb]
\begin{center}
\includegraphics*[width=7cm]{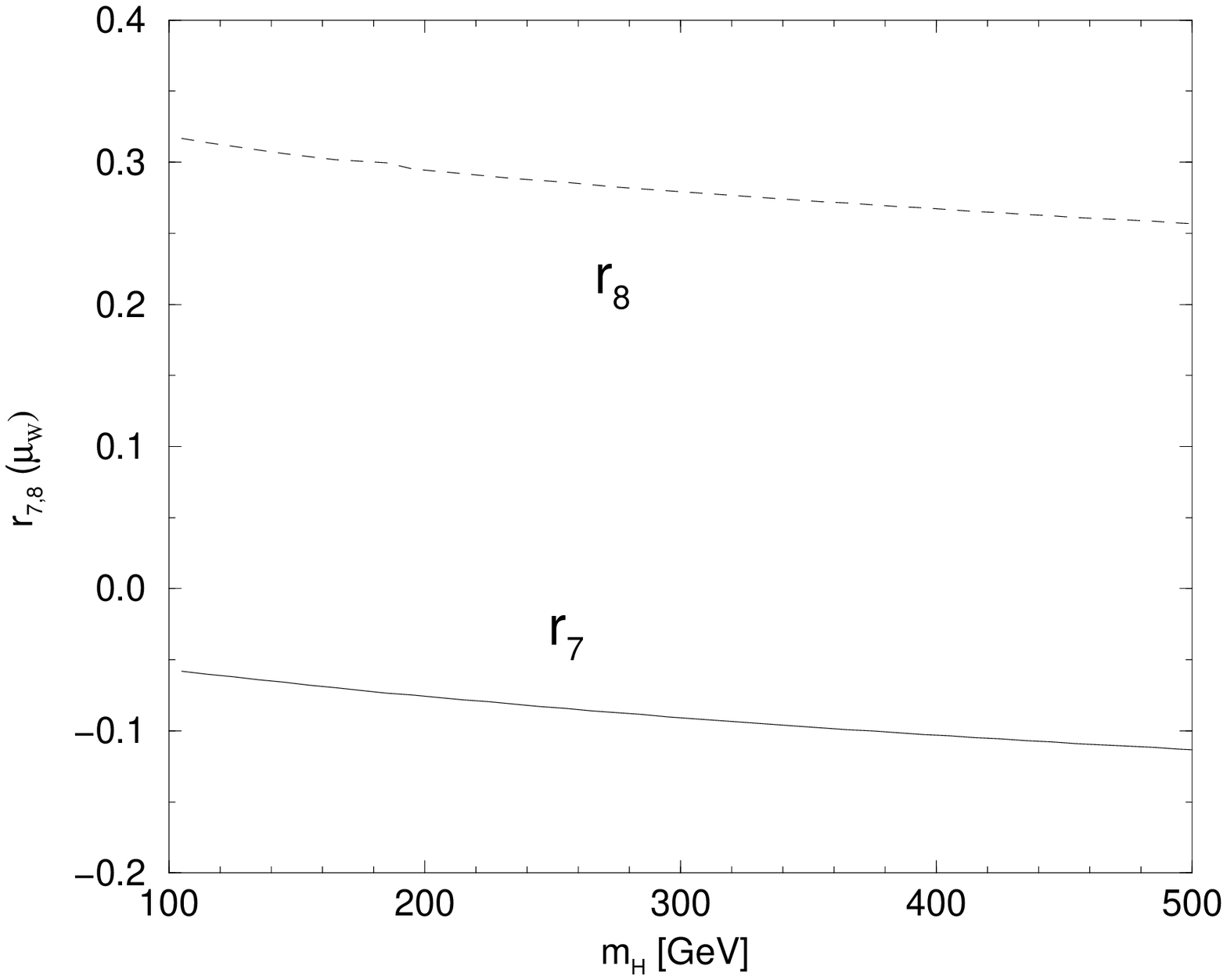}
\hspace{4mm}
\includegraphics*[width=7cm]{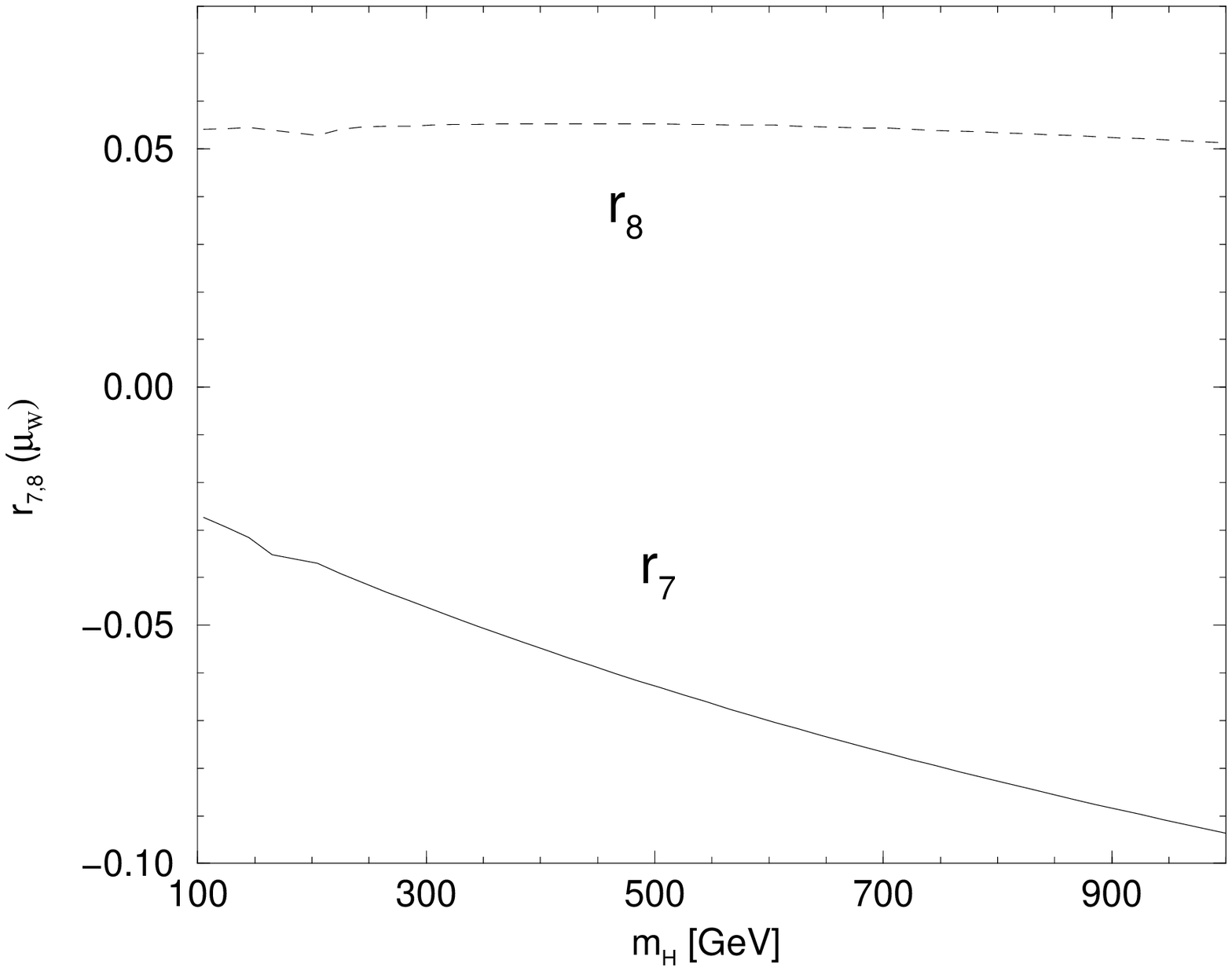}
\caption{%
Relative difference $r_i(\mu_W)(i=7,8)$ between the 
exact two-loop results and 
the nondecoupling approximations of $C_{i,H}(\mu_W)$, for 
the SUSY spectrum as in Fig.~\ref{figb1} (left) and heavier spectrum (right). 
}
\label{figb2}
\end{center}
\end{figure}

We perform an exact evaluation of the two-loop diagrams in 
Fig.~\ref{CHexch2loops} 
and compare the results to those in the nondecoupling approximation. 
In Fig.~\ref{figb1}, we show the numerical 
results of $C_{i,H}(\mu_W)$ as functions of $m_{H^{\pm}}$, 
for a SUSY particle spectrum 
$(m_{\tilde{s}_L},M_{\tilde{Q}_3},M_{\tilde{U}_3},M_{\tilde{D}_3})$ 
$=(250,230,210,260)$ GeV, $A_t=70$ GeV, $A_b=0$, 
$\tan\beta=30$, $m_{\tilde{g}}=200$ GeV, and $\mu=250$ GeV. 
We see that the $O(\alpha_s\tan\beta)$ corrections are numerically 
comparable to the one-loop results and must be included 
in realistic analysis. The deviation of the exact two-loop results 
is $O(m_{\rm weak}^2/M_{\rm SUSY}^2)$, the same order as 
the SU(2)$\times$U(1) breaking effects in the squark-gluino 
subloops~\cite{Burashuge}, and not negligible, 
especially for $C_{8,H}$. However, contrary to the 
naive expectation, the deviation does not show 
significant increase for $m_{H^{\pm}} > M_{\rm SUSY}$. This is 
more clearly seen in the left plot of Fig.~\ref{figb2} where 
the relative difference between the exact 
two-loop result and the nondecoupling approximation, 
\begin{equation}
r_i(\mu_W) \equiv 
  \frac{ C_{i,H}(\mu_W)\vert_{\rm nondec} -C_{i,H}(\mu_W)\vert_{\rm exact} } 
{ C_{i,H}(\mu_W)\vert_{\rm exact} } \hspace*{5truemm} (i = 7,8) , 
\end{equation}
is shown. For reference, the right plot of Fig.~\ref{figb2} 
shows the results for a heavier SUSY spectrum 
$(m_{\tilde{s}_L},M_{\tilde{Q}_3},M_{\tilde{U}_3},M_{\tilde{D}_3})=
(700,450,435,470)\,$GeV, $A_t = 150\,$GeV, $A_b=0$, 
$\tan\beta=30$, $m_{\tilde{g}}=600\,$GeV, and $\mu=550\,$GeV. 
$r_i$ is very small in the whole range of $m_{H^{\pm}}$. 
In both cases, the main part of the deviation comes from the 
diagram in Fig.~\ref{CHexch2loops}a) and, for $C_{8,H}$, also 
from the diagram in Fig.~\ref{CHexch2loops}e). 

To understand this unexpected result for $m_{H^\pm}>M_{\rm SUSY}$ 
qualitatively, we consider the diagram (a) in Fig.~\ref{CHexch2loops}, 
with chirality flip on the top quark line.  
When $m_{H^\pm}$ is sufficiently larger than $m_t$, 
this diagram gives the largest contribution 
to $\Delta C_{i,H}^1(\mu_W)$. The contribution is 
proportional to the loop integral 
\begin{equation}
 \mu m_{\tilde{g}} 
 I_{ti2}(m_t,m_{H^\pm}, m_{\tilde{t}_i},m_{\tilde{s}},m_{\tilde{g}}) =
\int\frac{d^4 k}{(2 \pi)^4} \
\frac{k^2}
{\left[k^2 -m_t^2\right]^3 \left[k^2 - m_{H^\pm}^2\right]}
 \,
 Y_{ti2}\left(k^2;m_{\tilde{t}_i},m_{\tilde{s}},m_{\tilde{g}}
       \right),  
\label{eq-Iint}
\end{equation}
where $Y_{ti2}(k^2;m_{\tilde{t}_i},m_{\tilde{s}},m_{\tilde{g}})$
represents the squark-gluino subdiagram contribution to the 
effective vertex $H^-\bar{s}_Lt_R$ and is given by 
\begin{equation}
Y_{ti2}(k^2;m_{\tilde{t}_i},m_{\tilde{s}},m_{\tilde{g}}) =
\mu m_{\tilde{g}} 
 \left[-2 F + (k^2-m_t^2) G\right]
\left(k^2;m_{\tilde{t}_i}^2,m_{\tilde{s}}^2,m_{\tilde{g}}^2\right),
\label{eq-Ydef} 
\end{equation}
with 
\begin{eqnarray}
F(k^2; m_{\tilde{t}_i}^2,m_{\tilde{s}}^2,m_{\tilde{g}}^2)
 & = & 
\int\frac{d^4 l}{(2 \pi)^4} \
\frac{1}{ \left[ (l+k)^2 - m_{\tilde{t}_i}^2 \right]
\left[ l^2 -\!m_{\tilde{s}}^2 \right]
\left[ l^2-m_{\tilde{g}}^2 \right] },
\label{eq-Fdef}
\\
k^{\mu} 
G(k^2; m_{\tilde{t}_i}^2,m_{\tilde{s}}^2,m_{\tilde{g}}^2) 
 & = & 
\int\frac{d^4 l}{(2 \pi)^4} \
\frac{l^{\mu}}{ \left[ (l+k)^2 - m_{\tilde{t}_i}^2 \right]
\left[ l^2 -\!m_{\tilde{s}}^2 \right]
\left[ l^2-m_{\tilde{g}}^2 \right]^2}.
\label{eq-Gdef}
\end{eqnarray}

In the nondecoupling approximation, the form factor 
$Y_{ti2}(k^2; m_{\tilde{t}_i},m_{\tilde{s}},m_{\tilde{g}})$ 
in Eq.~(\ref{eq-Iint}) is replaced by 
\begin{equation}
 Y_{ti2}\vert_{\rm nondec} =
 - 2 \mu m_{\tilde{g}} 
  F (0; m_{\tilde{t}_i}^2,m_{\tilde{s}}^2,m_{\tilde{g}}^2),
\label{eq-Yapprox}
\end{equation}
which is independent of $k^2$.  
For simplicity, we 
hereafter set $m_{\tilde{t}_i}$, $m_{\tilde{s}}$, $m_{\tilde{g}}$, and
$\mu$ equal to $M_{\rm SUSY}$.

For $|k^2|$ much smaller or larger than $M_{\rm SUSY}^2$, 
$Y_{ti2}(k^2; M_{\rm SUSY}^2)$ behaves as  
\begin{equation}
Y_{ti2}(k^2; M_{\rm SUSY}^2)  \to  
\left\{ 
\begin{array}{ll} 
 Y_{ti2}\vert_{\rm nondec} + 
 O\left(
 \displaystyle{\frac{k^2}{M_{\rm SUSY}^2}},
 \displaystyle{\frac{m_t^2}{M_{\rm SUSY}^2}} \right)  
& 
(\vert k^2 \vert \ll M_{\rm SUSY}^2), 
\\[1.8ex] 
 O\left(\displaystyle{\frac{M_{\rm SUSY}^2}{k^2} 
 \ln \frac{k^2}{M_{\rm SUSY}^2}} \right) 
& 
(\vert k^2 \vert \gg M_{\rm SUSY}^2), 
\end{array} 
\right. 
\label{eq-deviation}
\end{equation}
which supports the naive expectation that a substantial deviation of 
$I_{ti2}(m_t,m_{H^\pm},M_{\rm SUSY}^2)$ from 
$I_{ti2}(m_t,m_{H^\pm},M_{\rm SUSY}^2)\vert_{\rm nondec}$ may 
arise from the region $|k^2| > M_{\rm SUSY}^2$. 

However, the factor multiplying $Y_{ti2}(k^2; M_{\rm SUSY}^{2})$ in
Eqs.~(\ref{eq-Iint}) drops as $d^4k/k^6$ for $|k^2|\gg m_{H^\pm}^2$.
In fact, the bulk of the integral $I_{ti2}$ is determined by the 
small $|k^2|$ region up to $|k^2|=O(m_t^2)$. If $M_{\rm SUSY}$ is 
sufficiently larger than $m_t$, 
$Y_{ti2}(k^2; M_{\rm SUSY}^2)$ does not
deviate substantially from $Y_{ti2}\vert_{\rm nondec}$ in this region. 
This explains the smallness of the deviation for 
$m_{H^\pm}> M_{\rm SUSY}$ shown in Figs.~\ref{figb1} and \ref{figb2}. 

\begin{figure}[htb]
\begin{center}
\includegraphics[width= 6.0cm]{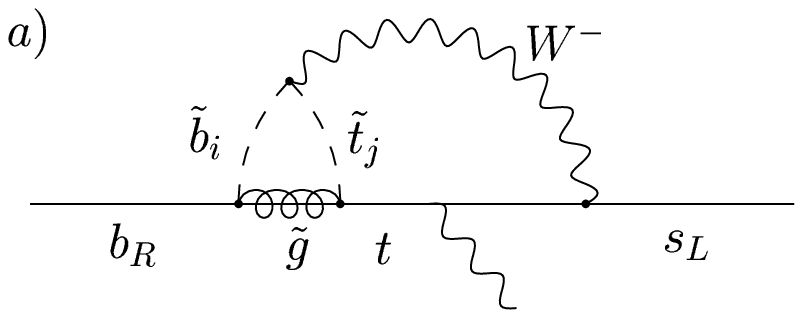}
\hspace{5mm}
\includegraphics[width= 6.0cm]{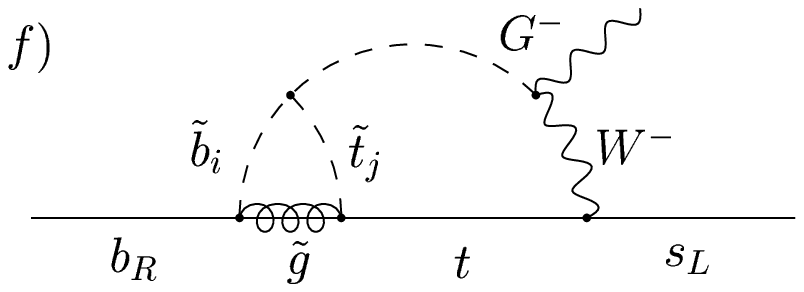}
\hspace{5mm}
\end{center}
\caption{%
Examples of the two-loop diagrams for the $O(\alpha\tan\beta)$ 
corrections to the $W^{\pm}$ contributions to $b\to s\gamma$. 
}
\label{figb4}
\end{figure}

We comment on the SUSY QCD corrections to other 
one-loop contributions to the $b\to(s\gamma, sg)$ decays. 
As already mentioned, the chargino contributions 
receive the $O(\alpha_s\tan\beta)$ correction 
to the $\tilde{\chi}^+b_R\tilde{t}^*_L$ 
coupling \cite{NLO-SUSY2a,NLO-SUSY2b} from $\delta m_b$ in 
Eq.~(\ref{eq:dmb}), through the coupling $h_b$ of the 
higgsino component $\widetilde{H}_D$ of $\tilde{\chi}^+$ to $b_R$. 
Other gluino 
corrections are not enhanced by $\tan\beta$ relative to the one-loop 
contribution. In contrast, the $W^{\pm}$ contributions 
do not receive $O(\alpha_s\tan\beta)$ corrections 
in the nondecoupling approximation. 
However, two-loop diagrams with effective $W^+\bar{t}b_R$ 
or $G^+\bar{t}b_R$ couplings, some of which are shown 
in Fig.~\ref{figb4}, give decoupling 
$O(\alpha_s\tan\beta \, m_{\rm weak}^2/M_{\rm SUSY}^2)$ 
contributions and may become nonnegligible for 
light $M_{\rm SUSY}\sim m_{\rm weak}$. Numerical study of 
these contributions 
will be presented in Ref.~\cite{BGYfuture}. 

In addition, there are also contributions to $b\to s\gamma$ coming from 
the mixings of squarks $\tilde{b}$ and $\tilde{s}$, such as 
the one-loop squark-gluino 
contribution \cite{BSGcontributions,bsglargeTB1,gluino}. 
One should note that squark generation mixings may be induced by 
the running of squark mass parameters \cite{squarkrunning}, 
$O(\tan\beta)$ corrections to the quark Yukawa 
matrices \cite{bsggluino1}, corrections to the 
squark-($\gamma,g$) couplings \cite{bsggluino2}, and 
other loop corrections. 
Studies of such contributions need consistent treatment of 
the squark sector renormalization including generation mixings, 
similar to the discussion in Section 2. 

\section{Conclusion}

We have discussed some aspects of 
the radiative corrections in the MSSM phenomenology, 
using two recent studies. 
First, the full one-loop corrections to the Higgs boson decays 
into charginos were presented. Especially, 
the renormalization of the chargino sector, 
including their mixing matrices, was discussed in detail. 
Numerical result was shown for the 
$(A^0,H^0)\to\tilde{\chi}^+_1\tilde{\chi}^-_1$ decays. 
Second, the two-loop $O(\alpha_s\tan\beta)$ corrections to 
the $b\to s\gamma$ and $b\to sg$ decays were discussed in models with 
large $\tan\beta$. Validity of the nondecoupling approximation, 
used in previous calculations, was examined for the $H^{\pm}$ contribution, 
by exact evaluation of the two-loop diagrams. 
The deviation was shown to be $O(m_{\rm weak}^2/M_{\rm SUSY}^2)$, 
but, contrary to naive expectation, not increase 
as $m_{H^\pm}$ even for $m_{H^{\pm}}> M_{\rm SUSY}$. 
A qualitative explanation for this unexpected behavior 
was presented in terms of the structure of the 
relevant two-loop integral.

\section{Acknowledgements}

I thank Helmut Eberl, Walter Majerotto, Francesca Borzumati, and 
Christoph Greub for fruitful collaborations on which this talk is 
based on. 
This work was supported by the 
Grant-in-aid for Scientific Research from the Ministry of Education,
Culture, Sports, Science, and Technology of Japan, No.~14740144.

\bibliographystyle{plain}

\end{document}